\documentclass[final,nofootinbib,aps,prl,twocolumn,showpacs,superscriptaddress,preprintnumbers,floatfix]{revtex4-1}

\usepackage{dynlearn}

\begin{document}

\def\ourTitle{  Optimized Bacteria are Environmental Prediction Engines
}

\def\ourAbstract{Experimentalists have observed phenotypic variability in isogenic bacteria
populations. We explore the hypothesis that in fluctuating environments this
variability is tuned to maximize a bacterium's expected log growth rate,
potentially aided by epigenetic markers that store information about past
environments. We show that, in a complex, memoryful environment, the maximal
expected log growth rate is linear in the instantaneous predictive
information---the mutual information between a bacterium's epigenetic markers
and future environmental states. Hence, under resource constraints, optimal
epigenetic markers are causal states---the minimal sufficient statistics for
prediction. This is the minimal amount of information about the past needed to predict the future as well as possible.  We suggest new theoretical
investigations into and new experiments on bacteria phenotypic bet-hedging in
fluctuating complex environments.
}

\def\ourKeywords{  epsilon-machine, causal states, predictive information, phenotypic bet-hedging
}

\hypersetup{
  pdfauthor={James P. Crutchfield},
  pdftitle={\ourTitle},
  pdfsubject={\ourAbstract},
  pdfkeywords={\ourKeywords},
  pdfproducer={},
  pdfcreator={}
}

\author{Sarah E. Marzen}
\email{semarzen@mit.edu}
\affiliation{Department of Physics,\\
Physics of Living Systems Group,\\
Massachusetts Institute of Technology, Cambridge, MA 02139}

\author{James P. Crutchfield}
\email{chaos@ucdavis.edu}
\affiliation{Complexity Sciences Center and Department of Physics, University of
  California at Davis, One Shields Avenue, Davis, CA 95616}

\date{\today}
\bibliographystyle{unsrt}

\title{\ourTitle}

\begin{abstract}
\ourAbstract
\end{abstract}

\keywords{\ourKeywords}

\pacs{
02.50.-r  89.70.+c  05.45.Tp  02.50.Ey  02.50.Ga  }

\preprint{\arxiv{1802.XXXX [cond-mat.stat-mech]}}

\title{\ourTitle}
\date{\today}
\maketitle

\setstretch{1.1}

\newcommand{\Abet}{\ProcessAlphabet}
\newcommand{\SSet}{\CausalStateSet}
\newcommand{\St}{\CausalState}
\newcommand{\st}{\causalstate}
\newcommand{\FSt}{\FutureCausalState}
\newcommand{\fst}{\futurecausalstate}
\newcommand{\FCmu}{\FutureCmu}
\newcommand{\PCmu}{\PastCmu}
\newcommand{\PSt}{\PastCausalState}
\newcommand{\pst}{\pastcausalstate}
\newcommand{\MxSt}{\AlternateState}
\newcommand{\MxSSet}{\AlternateStateSet}
\newcommand{\mxst}{\mu}
\newcommand{\mxstt}[1]{\mu_{#1}}
\newcommand{\StartMS}{\bra{\delta_\pi}}

\newcommand{\CodeRate}  { \I {\Past;\AlternateState} }
\newcommand{\Shielding} { \I {\Past;\Future | \AlternateState} }
\newcommand{\StateFutI} { \I {\AlternateState | \Future } }

\newcommand{\rep}    { y }
\newcommand{\Rep}    { Y }
\newcommand{\repa}    { \widehat{y} }
\newcommand{\Repa}    { \widehat{\mathcal{Y}} }
\newcommand{\repb}    { \widetilde{y} }
\newcommand{\Repb}    { \widetilde{\mathcal{Y}} }

Isogenic bacteria populations exhibit phenotypic variability
\cite{bonifield2003flagellar, moxon1994adaptive, bayliss2001simple,
kearns2004genes}. Some variability is unavoidable due to noise in the
underlying biological circuits and when and how they emerge during development
\cite{kaern2005stochasticity}. Such noise is not always detrimental to organism
functioning: phenotypic variability can be tuned to maximize population fitness
\cite{beaumont2009experimental, kussell2005phenotypic}. Such optimal phenotypic
variability is called \emph{bet hedging} \cite{Sege87a,de2011bet} and has been
implicated in seed germination in annual plants \cite{Cohe66a,Bulm84a} and in
phenotype switching by bacteriophages \cite{Masl15a} and fungi
\cite{soll1988comparison, jain2006phenotypic, guerrero2006phenotypic,
perez1999phenotypic}.

At first blush, it may seem strange that a population of organisms should not
simply express the phenotype that grows best in the most probable
environment---a deterministic strategy. Imagine, however, that the environment
fluctuates somewhat unpredictably (as real environments often do), sometimes
reaching a less probable state in which that phenotype does not reproduce. If
organisms only express that single phenotype, then eventually, the population
will go extinct. A population of organisms should, instead, hedge its ``bets''
about future environmental states, using the unavoidable noise in biological
circuits \cite{kaern2005stochasticity} or other mechanisms---e.g.,
slipped-strand mispairing \cite{moxon1994adaptive, bayliss2001simple}---to
express different phenotypes with varying probabilities. Given this, the only
question is: how should the population hedge its bets?

The first theoretical analysis of such bet-hedging was provided by Kelly in a
classic analysis of gambling; see Refs. \cite{kelly2011new} and \cite[Ch.
6]{Cove06a}. If one thinks of organisms as money, to draw out the parallel,
then gambling and bacterial growth are analogous. Adapting Kelly's setup, only
one phenotype can reproduce in any given environmental state. Kelly found in
effect that (i) the optimal probability of expressing a phenotype is the
probability of observing the corresponding environmental state and (ii) the
maximal expected log growth rate is linear in the negative entropy of a single
environmental state's probability.

Realistically, though, more than one phenotype might reproduce in a particular
environment. For example, a bacteria phenotype optimized for growth on a high
concentration of lactose can still grow on glucose, albeit with additional
energetic expenditure \cite{dekel2005optimality}. References
\cite{barron1988bound,bergstrom2004shannon} analyzed bet-hedging in just such
a case.

Furthermore, epigenetics provides a mechanism by which organisms can remember
the environmental past \cite{henikoff2016epigenetics}. This memory acts as
\emph{side information} about future environmental states---information that
can be used to increase the population's expected log growth rate
\cite{kelly2011new,Cove06a}.\footnote{In a different context, this observation
about memory was used to improve estimates of the entropy of written English
\cite{cover1978convergent}.} And, this suggests in turn that such memory should
affect optimal phenotypic variability. In fact, in the context of seed
germination, predictive cues about the current environmental state were found
to change the optimal germination fraction \cite{venable1980delayed}.

Here, we solve for optimal phenotypic variability and use this to calculate a
population's maximal expected log growth rate when accounting for both nonzero
reproduction rates of suboptimal phenotypes and epigenetic memory. We find that
the instantaneous predictive information---that shared between the organism's
present phenotype and future environment states---captures (and not just upper
bounds \cite{barron1988bound, bergstrom2004shannon}) the benefit of epigenetic
memory. When combined with resource constraints, this predicts that optimal
isogenic bacteria populations store epigenetic memories that are causal states
of bacteria observations of the environment. We conclude with suggestions
for testing and extending these results.

\paragraph*{Background}

Take the environment to be everything, except the bacteria phenotype, that
determines reproductive rates of an individual bacterium. At time $t$ the
environment is in a state $\present{t}$. What the bacteria observe of the
environment at time $t$ is $\obspresent{t}$---a noisy subsampling of the full
environmental state $\present{t}$ at time $t$. For example, the environmental
state $\present{t}$ might consist of a full list of available nutrients, only
some of which $\obspresent{t}$ are sensed by bacteria.

\begin{figure}
\centering
\includegraphics[width=0.45\textwidth]{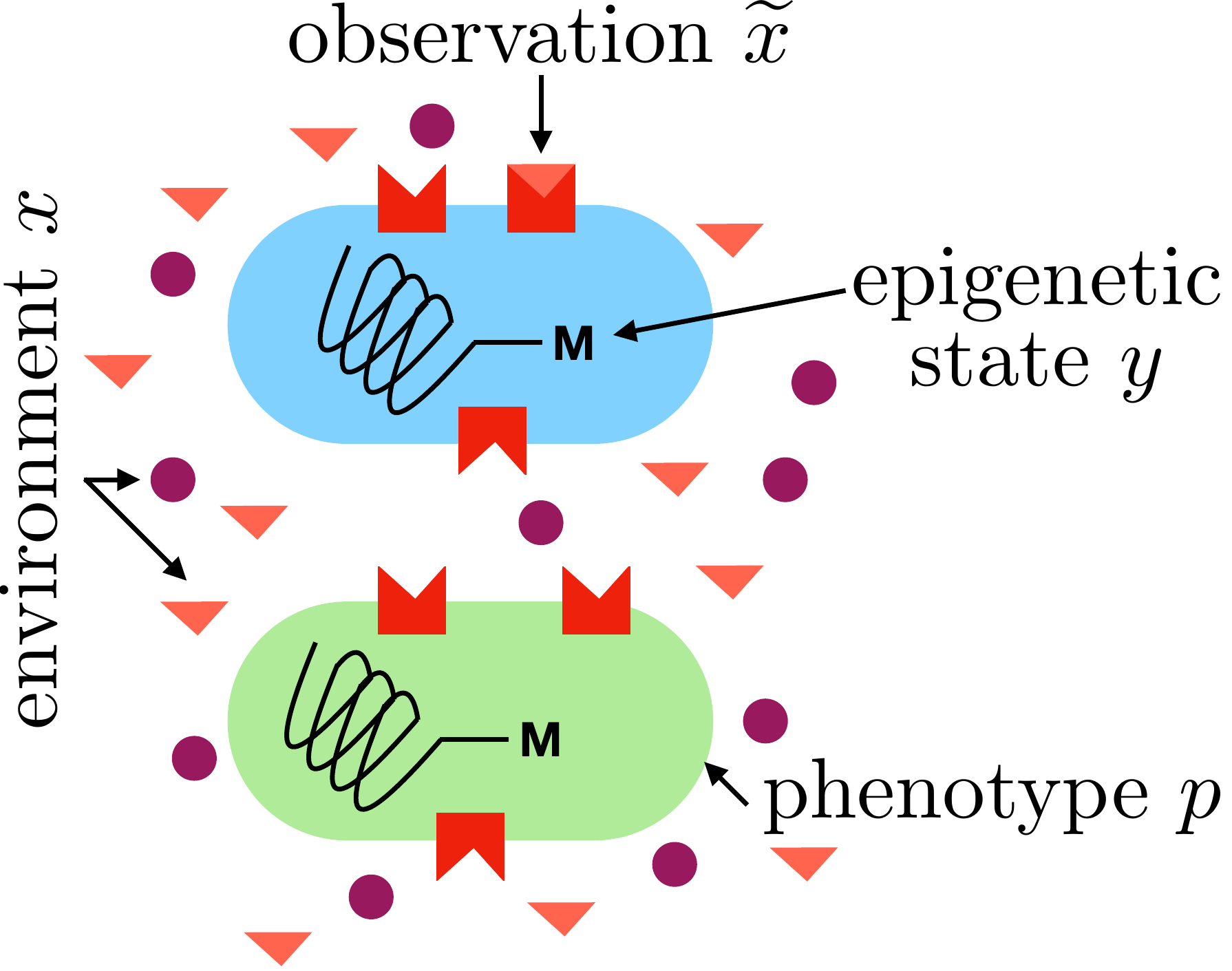}
\caption{A population of isogenic bacteria interacts with a fluctuating
	environment $x$, given by the concentrations of nutrients (pink and purple
	tokens). Bacteria observe only the concentration $\widetilde{x}$ of pink
	nutrient, remember aspects of the past environment through their epigenetic
	state $y$, which for instance could include their genome and any
	methylations $M$ thereof, and express a phenotype $p$ (blue or green ovals)
	that reproduces at different rates depending on the environment. The
	bacteria are isogenic, so that the epigenetic states of the two bacteria
	are identical. However, inherent biochemical stochasticity can cause the
	expression of different phenotypes.  Observations of the environment are
	assumed to be identical from bacterium to bacterium.
  }
\label{fig:setup}
\end{figure}

An individual bacterium has a genotype, an \emph{epigenetic state}---all the
epigenetic factors such as methylation that can be inherited---and a phenotype.
When we wish to emphasize that the epigenetic state contains information about
past environments, we refer to the state as an \emph{epigenetic memory}. We
denote the epigenetic state at time $t$ by $\rep_t$, with $\Rep_t$ its random
variable. See Fig. \ref{fig:setup}.

We assume the environmental time series $\ms{-\infty}{\infty} =
\ldots,\present{t},~\present{t+1},~\ldots$ is a realization of a stationary
stochastic process.  Time increments when the environment changes.  Bacteria
are assumed to stochastically choose a new phenotype every time step based on
their current epigenetic state. We assume that when a bacterium chooses its
phenotype, it only references its epigenetic state and not its previous
phenotype.

It is well worth mapping our setup's assumptions to those previously used to
explore the value of information for populations subject to fluctuating
environments \cite{rivoire2011value}. We simultaneously relax assumptions $A1$
(``no information is inherited'') and $A3$ (``only one phenotype survives'')
there, allowing for inheritance only through the epigenetic state, not through
the previous phenotype. The last assumption does not map onto any in Ref.
\cite{rivoire2011value}. It constitutes the main insight that allows relaxing
both A$1$ and A$3$ but still yields closed-form expression for the value of
information as the increase in expected log growth rate arising from storing
information about the environment. We do not allow each bacterium to observe
the environment differently; in other words, Ref. \cite{rivoire2011value}'s
environmental sensor $q(\rep_t|\present{t})$ is the identity map.

We later argue that a bacterium should optimally predict its environment,
(somehow) using the environment's causal states \cite{Shal98a}. Two observed
environmental pasts $\obs{-\infty}{t}$ and $\obs{-\infty}{t}'$ are considered
equivalent, $\obs{-\infty}{t}\sim_{\epsilon}\obs{-\infty}{t}'$, if and only if
$\Prob(\Obspresent{t}|\Obs{-\infty}{t}=\obs{-\infty}{t})=\Prob(\Obspresent{t}|\Obs{-\infty}{t}=\obs{-\infty}{t}')$.
In this, $\Obs{-\infty}{t} = \ldots \Obspresent{t-2}, \Obspresent{t-1}$ is the
chain of random variables representing the observed pasts. The equivalence
relation $\sim_{\epsilon}$ partitions the set of all pasts into classes called
\emph{causal states} $\sigma^+\in\SSet$ and induces a rule that maps a past to
its causal state: $\causalstate = \epsilon(\obs{-\infty}{t})$. Causal states
are the minimal sufficient statistics for predicting the environment, meaning
that they constitute the minimal information about the past necessary to
predict the future as well as one possibly could given the
observations.\footnote{The reinforcement learning literature has come to call
causal states \emph{predictive representations} \cite{littman2001predictive}.}

Let $\St_t$ be the random variable corresponding to the causal state at time
$t$. From the probabilities
$\Prob(\Obspresent{t}|\Obs{-\infty}{t})$ and the rule $\epsilon(\cdot)$, one
obtains a transition dynamic $\Prob(\St_{t+1},\Obspresent{t}|\St_t)$ on causal
states. The corresponding hidden Markov model is the environment's minimal,
optimal model---its \emph{\eM} \cite{Shal98a}. It is \emph{unifilar}---that is,
given the environment's current causal state and next observation, its next
state is uniquely determined. Of the unifilar hidden Markov models that
describe a given environment, the \eM\ has the minimal number of states \cite{Shal98a,Trav11a}.

\paragraph*{Results}
First, we obtain an expression for the expected log growth rate and maximize
this with respect to phenotypic variability. We find that the instantaneous
predictive information defines the quality of an epigenetic state under some
assumptions on reproduction rates and environmental statistics. Then, we show
that the optimal resource-constrained epigenetic states are the observational
causal states. Importantly, this latter result is free from some of the more stringent assumptions
required to establish the former result.

\newcommand{\pheno} {p}

\paragraph*{Emergence of instantaneous predictive information}
Let $n_t$ be the number of
organisms at time $t$. Let $\Pr(\pheno|\rep_t)$ be a bacterium's
\emph{strategy}---the probability that an
organism expresses phenotype $\pheno$ given epigenetic state $\rep_t$.  This
conditional probability distribution exists in a \emph{strategy simplex}---the space of valid conditional probability distributions $\Pr(\pheno|\rep)$.
Assume that a bacterium's phenotype at the next time step depends on the epigenetic state but is generated
independently of its phenotype at the previous time step. Finally, let
$f(\pheno,\present{})$ be the reproduction rate of phenotype $\pheno$ in
environment $\present{}$, which might depend on the energetic efficiency of
that phenotype in that environment.

Then, we straightforwardly obtain:
\begin{align*}
n_{t+1} & = \sum_{\pheno} \left( \Pr(\pheno|\rep_t) n_t \right)
	f(\pheno,\present{t}) \\
        & = \left(\sum_{\pheno} \Pr(\pheno|\rep_t) f(\pheno,\present{t}) \right) n_t
  ~.
\end{align*}
This yields an expected log-growth rate:
\begin{align}
r & = \left\langle \log\frac{n_{t+1}}{n_t} \right\rangle \nonumber \\
  & = \left\langle \log \left(\sum_{\pheno} \Pr(\pheno|\rep_t) f(\pheno,\present{t}) \right)
  	\right\rangle \nonumber \\
  & = \sum_{\rep_t,\present{t}} \Pr(\rep_t,\present{t})
  \log \left(\sum_{\pheno} \Pr(\pheno|\rep_t) f(\pheno,\present{t}) \right) \label{eq:r}
  ~.
\end{align}
We focus on the expected log-growth rate as a natural measure of population
fitness rather than on the fitness of an individual, which might be better
quantified by expected growth rate. Why? In the case of phenotypic bet-hedging,
what is good for the population is not necessarily good for the individual. To
survive, an individual should choose a strategy that survives in all
environments, even if it grows slowly in some. A population, however, has the
luxury of having some organisms bet on phenotypes that might not survive in
certain environments if they grow much faster in others.  Hence, we are
interested in what kinds of isogenic bacterial \emph{populations} evolve.
However, since these populations are isogenic, we describe the evolved
population in terms of properties of the individual bacterium.

Also, note that $\rep_t$ has access to information about $\obs{-\infty}{t}$ but
cannot directly access information about $\ms{-\infty}{t}$. All of $\rep_t$'s
information about $\present{t}$ comes through $\obs{-\infty}{t}$; i.e., we have
the Markov chain $\Rep_t \rightarrow \Obs{-\infty}{t} \rightarrow
\MS{-\infty}{t} \rightarrow \Present{t}$.

We seek the bet-hedging strategy $\Pr(\pheno|\rep_t)$ that maximizes expected
log-growth rate $r$. Our derivation closely follows that of Ref.
\cite{bergstrom2004shannon}, with the key change that we now allow for
side-information from epigenetic memory. We maximize $r$, subject to the
constraint that $\sum_g \Pr(\pheno|\rep_t) = 1$ for all $\rep_t$, via the
Lagrangian:
\begin{align*}
\mathcal{L} & = \sum_{\rep_t,\present{t}} \Pr(\rep_t,\present{t})
  \log \left(\sum_{\pheno} \Pr(\pheno|\rep_t) f(\pheno,\present{t}) \right) \\
  & \qquad + \sum_{\rep_t} \lambda_{\rep_t} \sum_g \Pr(\pheno|\rep_t)
  ~,
\end{align*}
with respect to $\Pr(\pheno|\rep_t)$, where $\lambda_{\rep_t}$ is the Lagrange
multiplier for each epigenetic state $\rep_t$. Note that if the bacteria
population strongly affected the environment's dynamics, then
$\Pr(\rep_t,\present{t})$ would depend on $\Pr(\pheno|\rep_t)$. Instead, we
assume the environment is so large that the bacteria population does not
affect it.

\newcommand{\One}{ {\mathbf{1}} } 

To find the strategy $\Pr(\pheno|\rep_t)$ that maximizes $r$, we take
derivatives of the Lagrangian and set them to $0$:
\begin{align*}
0 & = \frac{\partial \mathcal{L}}{\partial \Pr(\pheno|\rep)} \\
  & = \sum_{\present{}} \Pr(\present{}|\rep)
  \frac{f(\pheno,\present{})}
  {\sum_{\pheno} \Pr(\pheno|\rep) f(\pheno,\present{})}
  - \lambda_{\rep}
  ~.
\end{align*}
And so:
\begin{align*}
\lambda_{\rep} &= \sum_{\present{}} \Pr(\present{}|\rep)
  \frac{f(\pheno,\present{})}{\sum_{\pheno} \Pr(\pheno|\rep) f(\pheno,\present{})}
  ~.
\end{align*}

Let $\bf{x}_{\rep}$ be the vector of optimal strategies $\Pr(\pheno|\rep)$,
$\bf{p}_{\rep}$ the vector of $\Pr(\present{t}|\rep)$, and $W$ the matrix with
elements $f(\pheno,\present{})$. Then, the preceding result in matrix form is:
\begin{align*}
\lambda_{\rep} \One &= W \left({\bf p}_{\rep} \odot [W^{\top} {\bf x}_{\rep}]^{\odot -1}\right)
  ~,
\end{align*}
where the $1$s vector $\One$ has the length of the number of possible
phenotypes and $\odot$ is the Hadamard product, so that $\odot$ represents
componentwise multiplication and $[W^{\top} {\bf x}_{\rep}]^{\odot -1}$
represents componentwise inversion. If $W$ is invertible, then we solve for
$\bf{x}_{\rep}$:
\begin{align*}
{\bf x}_{\rep} & = \frac{1}{\lambda_{\rep}}
   \left( W^{\top} \right)^{-1}
   \left({\bf p}_{\rep}\odot [W^{-1}\One]^{\odot -1}\right)
  ~,
\end{align*}
and, using the normalization condition $\One^{\top} \bf{x}_{\rep} = 1$, we
fortuitously find that:
\begin{align}
{\bf x}_{\rep} &= \left(W^{\top} \right)^{-1}
   \left({\bf p}_{\rep}\odot [W^{-1}\One]^{\odot -1}\right). \label{eq:optx}
\end{align}
Note that this is the maximizing conditional distribution \emph{if} it is in
the strategy simplex and \emph{if} $W$ is invertible. One might relax the
condition that $W$ is invertible, if $W$ is square, via the Drazin inverse
\cite[Sec. IV.H]{Riec16a}. In sum, Eq. (\ref{eq:optx}) gives the optimal
strategy for phenotypic variability given a particular epigenetic memory.

Recall from Eq. (\ref{eq:r}) that the expected log-growth rate $r$ is a
function of epigenetic memories $\rep_t$ via the average over
$\Prob(\rep_t,\present{t})$, the phenotypic strategy $\Prob(\pheno|\rep_t)$,
and reproductive rates $f(\pheno,\present{t})$. Given the optimal strategy
${\bf x}_{\rep}$ from Eq. (\ref{eq:optx}), one finds a maximal expected
log-growth rate:
\begin{align}
r^* & = \sum_{\rep_t,\present{t}}
  \Pr(\rep_t,\present{t})
  \log \frac{\Pr(\present{t}|\rep_t)}{\sum_{\pheno} (W^{-1})_{\pheno,\present{t}}} \nonumber \\
  & = -\H{\Present{t}|\Rep_t}
  - \sum_{\present{t}} \Pr(\present{t}) \log \sum_{\pheno}
	(W^{-1})_{\pheno,\present{t}}
  ~. \label{eq:finalr}
\end{align}
The first $-\H{\Present{t}|\Rep_t}$ of these two terms depends on the scheme
that associates epigenetic states to environmental pasts. The second is
independent of such schemes and depends only on environmental statistics and
reproduction rates.

Now, recall that Ref.  \cite{rivoire2011value}'s ``value of information'' $\Delta r^*$ is
the increase in maximal expected log-growth rate of a population with epigenetic
memory above and beyond that of a population without any epigenetic memory. And so, if Eq. (\ref{eq:optx}) yields an $\bf{x}_{\rep}$ in the
strategy simplex, then the ``value of
information'' is:
\begin{align}
\Delta r^*
  & = -\H{\Present{t}|\Rep_t} + \H{\Present{t}} \nonumber \\
  & = \I{\Rep_t;\Present{t}}
  ~.
\end{align}
This is the instantaneous predictive information \cite{Bial00a,Still2012}.
(Note the difference in notation between here and Ref. \cite{Still2012}, in
that here, $\rep_t$ lags behind $\present{t}$ by a half-time step.) Hence,
epigenetic states with higher instantaneous predictive information are
evolutionarily favored.

\paragraph*{Optimal epigenetic states are causal states}
Earlier, we stated that Eq. (\ref{eq:optx}) gave the optimal phenotypic
variability for a given epigenetic memory when the associated strategy was in
the strategy simplex.  If so, then the Data Processing Inequality
\cite{Cove06a} reveals that:
\begin{align}
\I{\Rep_t;\Present{t}} \leq \I{\Obs{-\infty}{t};\Present{t}} \leq
\I{\MS{-\infty}{t};\Present{t}}
  ~.
\label{eq:ineq}
\end{align}
Employing the Data Processing Inequality, we implicitly assume that a
bacterium's only guide to the future environment consists of past environmental
states. In other words, we assume that an experimentalist, say, does not give
the bacterium \emph{additional} side information about the environment. The
quantity $\I{\MS{-\infty}{t};\Present{t}} = \H{\Present{t}} - \hmu$ is also
known as the \emph{predicted information rate} or the \emph{total correlation
rate} \cite{Jame11a,Jame13a}. It is largely controlled by the environment's
intrinsic randomness or Shannon entropy rate $\hmu =
\H{\Present{t}|\MS{-\infty}{t}}$.

Equation (\ref{eq:ineq}) suggests evolution favors populations of organisms
that develop epigenetic memories which as much of the environmental past as
possible. However, memory is costly and one should not remember environmental
pasts that are not helpful. More specifically, genomes are finite in size and
can only support a finite number of epigenetic markers. Hence, the number
$|\mathcal{Y}|$ of possible epigenetic states is finite. The balance to strike
therefore is to saturate the inequality in Eq. (\ref{eq:ineq}) while minimizing
a resource cost---the number $|\mathcal{Y}|$ of possible epigenetic states. In
short, epigenetic memories store the minimal amount of information about the
observed environment's past needed to predict the environment's future. They
are, therefore, the minimal sufficient statistics of prediction of the future
environment with respect to past observations.

How might epigenetic memories store such information? After all, at a given
time $t$ a bacterium cannot directly access the observed environment's past
$\obs{-\infty}{t}$. However, a bacterium's future epigenetic state $\rep_{t+1}$
depends on both its previous epigenetic state $\rep_t$ and the present
environmental observation $\present{t}$. In other words, a bacterium's
epigenetic state is generated by an input-dependent dynamical system whose
input is the environmental observation. If the update rule for how the
bacterium's future epigenetic state $\rep_{t+1}$ depends on the previous
epigenetic state $\rep_t$ and the present environmental observation
$\present{t}$ are chosen so as to mimic the environment's \eM\ transition
dynamic, then the bacterium's epigenetic state $\rep_t$ at time $t$ will be the
environment's causal state \cite{Shal98a}. This is the limit to what is
realizable from an input-dependent dynamical system.  Hence, a bacterium's
optimal \emph{realizable} epigenetic memories are causal states of the observed
environment.

More generally, Eq. (\ref{eq:optx}) might not give a valid conditional
probability distribution or the matrix $W$ there might not be invertible. Even
then, maximization of expected log growth rate combined with resource
limitations implies that optimal epigenetic memories are causal states. To show
this, we first show that expected log growth rate is maximized when the
epigenetic memories store the entire observed environmental past. Then, we show that
this maximum is also achieved when epigenetic memories are minimal sufficient statistics of prediction
of the future environment with respect to past observations.
Finally, the aforementioned resource constraints imply that optimal realizable epigenetic
memories are causal states.

Let's explain this and so provide a sketch of its proof. As stated, we must
first show that expected log growth rate is maximized when the epigenetic
memories store the entire environmental past. To see this, note that any
$\Pr(\pheno|\rep)$, for any realizable $\rep$, can be represented if
$\rep_t=\obs{-\infty}{t}$. Hence:
\begin{align*}
\max_{\Pr(\pheno_t|\rep_t)} r \leq \max_{\Pr(\pheno_t|\obs{-\infty}{t})} r
  ~.
\end{align*}
Then, as desired:
\begin{align*}
\max_{\Pr(\rep_{t}|\obs{-\infty}{t})} \max_{\Pr(\pheno_t|\rep_t)} r =
\max_{\Pr(\pheno_t|\rep_t):\rep_t=\obs{-\infty}{t}} r
  ~.
\end{align*}
Next, we show that this maximum is also achieved when epigenetic memories are
minimal sufficient statistics of prediction
of the future environment with respect to past observations. Note that the expression for $r$ is linear in
$\Pr(\present{t},\rep_t)$, and so $\max_{\Pr(\pheno|\rep)} r$ depends only on
$\Pr(\present{t}|\rep_t)$, averaged over $\Pr(\rep_t)$. This, in turn, implies that maximal expected log
growth rate can be achieved by any sufficient statistic of prediction. If we prefer sufficient statistics with smaller
$|\mathcal{Y}|$, then we find that optimal realizable epigenetic memories are causal
states \cite{Shal98a}, as stated earlier.

\paragraph*{Conclusions}
We proposed that isogenic bacteria populations must predict their environment
to maximize their expected log growth rate. We justified this via extensions of
Kelly's classic bet-hedging analysis that follow Ref.
\cite{bergstrom2004shannon}. This conclusion and Eq. (\ref{eq:optx}) give
explicitly-testable predictions for new kinds of bacteria evolution experiment
in which populations evolve subject to a fluctuating \emph{memoryful}
environment. For instance, one can subject populations to partly-random,
partly-predictable patterns of antibiotics. The prediction is that the bacteria
will develop optimal phenotypic bet-hedging behavior in which their probability
of exhibiting a particular phenotype implies epigenetic memory; i.e., with
phenotypic variability given by Eq. (\ref{eq:optx}) and with epigenetic memories that correspond to causal states of the environment. Although the above
analysis focused on bacteria, similar results apply to the phenotype-switching
fungi cited earlier.

That said, Ref. \cite{kussell2005phenotypic}'s setup might be more appropriate
for interfacing with experiment. As such, we briefly describe an extension of
that setup that should yield similar qualitative results to those presented
here. Reference \cite{kussell2005phenotypic} studied phenotypic bet-hedging in
a continuous-time system and assessed the difference between stochastically
switching phenotypes (bet-hedging) and switching to the best phenotype based on
sensing. In point of fact, there is a time delay between sensing and action
that can be explicitly built into a model of environmental sensing and
phenotypic switching. One should then find that \emph{memory} of past
environmental states, above and beyond instantaneous sensing of present
environmental states, can be used to better select the next phenotype. The
environment's inherent stochasticity will also lead such optimally-sensing
populations to not only utilize memory of past fluctuations, but also to
stochastically choose phenotypes.

For randomly selected processes, their optimal predictors (\eMs) are usually
not finite. They can often be very large even when finite. Thus, the resource
constraints mentioned earlier become paramount when addressing more
naturalistic environments. It is surprisingly easy to put resource constraints
and predictive information on the same footing in this setup based solely on
their effect on the expected log-growth rate.

Consider Eq. (\ref{eq:finalr}). If there are more stringent constraints on
bacteria size, then reproductive rates $f(\pheno,\present{})$ might increase,
since less material is required to generate a new bacterium. Therefore,
resource constraints will increase the second term in Eq. (\ref{eq:finalr}).
However, stronger resource constraints tend to diminish the predictive
information captured by a bacteria population, as given by the first term in
Eq. (\ref{eq:finalr}). Hence, one expects the input-dependent dynamical system
supporting a bacterium's epigenetic states to find ``lossy causal states''
\cite{Marz14f} rather than causal states. In this, the degree of tradeoff
between resource constraints and predictive information is determined by the
environment and the organism's ability to grow in said environment. Lossy
causal states can be calculated using the methods of Ref. \cite{Marz14f}.

The derivation above assumed that the environment was so large that its
evolution was independent of bacteria phenotypes.  However, bacteria certainly
affect their environment, at the very least by secreting molecules and removing
nutrients. Ideally, we would not assume that the environment's evolution was
independent of the bacteria's actions, thereby closing the sensorimotor loop
and allowing for niche construction \cite{Odl03a}. We expect relaxing this
assumption to yield much more complicated quantifiers of the quality of
epigenetic memory, given the difficult of solution of partially observable
Markov decision processes (POMDPs); e.g., as described in Refs.
\cite{meuleau1999solving, doshi2015bayesian, hausknecht2015deep}. However, we
expect causal states to be optimal epigenetic states, since the belief states
used in the solution of POMDPs are causal states.

\section*{Acknowledgments}

The authors thank J. Horowitz, D. Amor, A. Solon, J. England, C. Ellison, C.
Hillar, and S. DeDeo for helpful discussions, and the Santa Fe Institute for
its hospitality during visits, where JPC is an External Faculty member. This
material is based upon work supported by, or in part by, the John Templeton
Foundation grant 52095, the Foundational Questions Institute grant
FQXi-RFP-1609, the U. S. Army Research Laboratory and the U. S. Army Research
Office under contract W911NF-13-1-0390. S.E.M. was funded by an MIT Physics of
Living Systems Fellowship.


\begin{thebibliography}{10}

\bibitem{bonifield2003flagellar}
H.~R. Bonifield and K.~T. Hughes.
\newblock Flagellar phase variation in \emph{Salmonella enterica} is mediated
  by a posttranscriptional control mechanism.
\newblock {\em J. Bacterio.}, 185(12):3567--3574, 2003.

\bibitem{moxon1994adaptive}
E.~R. Moxon, P.~B. Rainey, M.~A. Nowak, and R.~E. Lenski.
\newblock Adaptive evolution of highly mutable loci in pathogenic bacteria.
\newblock {\em Current Biology}, 4(1):24--33, 1994.

\bibitem{bayliss2001simple}
C.~D. Bayliss, D.~Field, and E.~R. Moxon.
\newblock The simple sequence contingency loci of \emph{Haemophilus influenzae}
  and \emph{Neisseria meningitidis}.
\newblock {\em J. Clinical Invest.}, 107(6):657, 2001.

\bibitem{kearns2004genes}
D.~B Kearns, F.~Chu, R.~Rudner, and R.~Losick.
\newblock Genes governing swarming in \emph{Bacillus subtilis} and evidence for
  a phase variation mechanism controlling surface motility.
\newblock {\em Molecular Microbio.}, 52(2):357--369, 2004.

\bibitem{kaern2005stochasticity}
M.~K{\ae}rn, T.~C Elston, W.~J. Blake, and J.~J. Collins.
\newblock Stochasticity in gene expression: {From} theories to phenotypes.
\newblock {\em Nature Rev. Gen.}, 6(6):451--464, 2005.

\bibitem{beaumont2009experimental}
H.~J.~E. Beaumont, J.~Gallie, C.~Kost, G.~C. Ferguson, and P.~B Rainey.
\newblock Experimental evolution of bet hedging.
\newblock {\em Nature}, 462(7269):90--93, 2009.

\bibitem{kussell2005phenotypic}
E.~Kussell and S.~Leibler.
\newblock Phenotypic diversity, population growth, and information in
  fluctuating environments.
\newblock {\em Science}, 309(5743):2075--2078, 2005.

\bibitem{Sege87a}
J.~Seger and H.~J. Brockmann.
\newblock What is bet-hedging?
\newblock {\em Oxford Surveys in Evolutionary Biology}, 4:182--211, 1987.

\bibitem{de2011bet}
I.~G. de~Jong, P.~Haccou, and O.~P. Kuipers.
\newblock Bet hedging or not? {A} guide to proper classification of microbial
  survival strategies.
\newblock {\em Bioessays}, 33(3):215--223, 2011.

\bibitem{Cohe66a}
D.~Cohen.
\newblock Optimizing reproduction in a randomly varying environment.
\newblock {\em J. Theo. Bio.}, 12:119--129, 1966.

\bibitem{Bulm84a}
M.~G. Bulmer.
\newblock Delayed germination of seeds: {Cohen's} model revisited.
\newblock {\em Theo. Pop. Bio.}, 26:367--377, 1984.

\bibitem{Masl15a}
S.~Maslov and K.~Sneppen.
\newblock Well-temperate phage: optimal bet-hedging against local environmental
  collapses.
\newblock {\em Sci. Reports}, 5:10523, 2015.

\bibitem{soll1988comparison}
D.~R. Soll and B.~Kraft.
\newblock A comparison of high frequency switching in the yeast \emph{Candida
  albicans} and the slime mold \emph{Dictyostelium discoideum}.
\newblock {\em Genesis}, 9(4-5):615--628, 1988.

\bibitem{jain2006phenotypic}
N.~Jain, A.~Guerrero, and B.~C. Fries.
\newblock Phenotypic switching and its implications for the pathogenesis of
  \emph{Cryptococcus neoformans}.
\newblock {\em FEMS Yeast Res.}, 6(4):480--488, 2006.

\bibitem{guerrero2006phenotypic}
A.~Guerrero, N.~Jain, D.~L. Goldman, and B.~C. Fries.
\newblock Phenotypic switching in cryptococcus neoformans.
\newblock {\em Microbiology}, 152(1):3--9, 2006.

\bibitem{perez1999phenotypic}
J.~P{\'e}rez-Mart{\'\i}n, J.~A. Ur{\'\i}a, and A.~D. Johnson.
\newblock Phenotypic switching in \emph{Candida albicans} is controlled by a
  {SIR2} gene.
\newblock {\em EMBO J.}, 18(9):2580--2592, 1999.

\bibitem{kelly2011new}
J.~L. Kelly.
\newblock A new interpretation of information rate.
\newblock In {\em The Kelly Capital Growth Investment Criterion: Theory and
  Practice}, pages 25--34. World Scientific, 2011.

\bibitem{Cove06a}
T.~M. Cover and J.~A. Thomas.
\newblock {\em Elements of Information Theory}.
\newblock Wiley-Interscience, New York, second edition, 2006.

\bibitem{dekel2005optimality}
E.~Dekel and U.~Alon.
\newblock Optimality and evolutionary tuning of the expression level of a
  protein.
\newblock {\em Nature}, 436(7050):588--592, 2005.

\bibitem{barron1988bound}
A.~R. Barron and T.~M. Cover.
\newblock A bound on the financial value of information.
\newblock {\em IEEE Trans. Info. Th.}, 34(5):1097--1100, 1988.

\bibitem{bergstrom2004shannon}
C.~T. Bergstrom and M.~Lachmann.
\newblock Shannon information and biological fitness.
\newblock In {\em Information Theory Workshop, 2004. IEEE}, volume IEEE
  0-7803-8720-1, pages 50--54, 2004.

\bibitem{henikoff2016epigenetics}
S.~Henikoff and J.~M. Greally.
\newblock Epigenetics, cellular memory and gene regulation.
\newblock {\em Current Biology}, 26(14):R644--R648, 2016.

\bibitem{cover1978convergent}
T.~Cover and R.~King.
\newblock A convergent gambling estimate of the entropy of {English}.
\newblock {\em IEEE Trans. Info. Th.}, 24(4):413--421, 1978.

\bibitem{venable1980delayed}
D.~L. Venable and L.~Lawlor.
\newblock Delayed germination and dispersal in desert annuals: {Escape} in
  space and time.
\newblock {\em Oecologia}, 46(2):272--282, 1980.

\bibitem{rivoire2011value}
O.~Rivoire and S.~Leibler.
\newblock The value of information for populations in varying environments.
\newblock {\em J Stat. Physics}, 142(6):1124--1166, 2011.

\bibitem{Shal98a}
C.~R. Shalizi and J.~P. Crutchfield.
\newblock Computational mechanics: Pattern and prediction, structure and
  simplicity.
\newblock {\em J. Stat. Phys.}, 104:817--879, 2001.

\bibitem{littman2001predictive}
M.~L. Littman, R.~S. Sutton, and S.~P. Singh.
\newblock Predictive representations of state.
\newblock In {\em NIPS}, volume~14, pages 1555--1561, 2001.

\bibitem{Trav11a}
N.~Travers and J.~P. Crutchfield.
\newblock Equivalence of history and generator $\epsilon$-machines.
\newblock arxiv.org:1111.4500.

\bibitem{Riec16a}
P.~M. Riechers and J.~P. Crutchfield.
\newblock Beyond the spectral theorem: Decomposing arbitrary functions of
  nondiagonalizable operators.
\newblock 2016.
\newblock Santa Fe Institute Working Paper 16-07-015; arxiv.org:1607.06526
  [math-ph].

\bibitem{Bial00a}
W.~Bialek, I.~Nemenman, and N.~Tishby.
\newblock Predictability, complexity, and learning.
\newblock {\em Neural Computation}, 13:2409--2463, 2001.

\bibitem{Still2012}
S.~Still, D.~A. Sivak, A.~J. Bell, and G.~E. Crooks.
\newblock Thermodynamics of prediction.
\newblock {\em Phys. Rev. Lett.}, 109:120604, Sep 2012.

\bibitem{Jame11a}
R.~G. James, C.~J. Ellison, and J.~P. Crutchfield.
\newblock Anatomy of a bit: {Information} in a time series observation.
\newblock {\em CHAOS}, 21(3):037109, 2011.

\bibitem{Jame13a}
R.~G. James, K.~Burke, and J.~P. Crutchfield.
\newblock Chaos forgets and remembers: {Measuring} information creation,
  destruction, and storage.
\newblock {\em Phys. Lett. A}, 378:2124--2127, 2014.

\bibitem{Marz14f}
S.~E. Marzen and J.~P. Crutchfield.
\newblock Predictive rate-distortion for infinite-order {Markov} processes.
\newblock {\em J. Stat. Phys.}, 163(6):1312--1338, 2016.

\bibitem{Odl03a}
F.~J. Odling-Smee, K.~N. Laland, and M.~W. Feldman.
\newblock {\em Niche Construction: The Neglected Process in Evolution}.
\newblock Princeton University Press, Princeton, New Jersey, 2003.

\bibitem{meuleau1999solving}
N.~Meuleau, K.-E. Kim, L.~P. Kaelbling, and A.~R. Cassandra.
\newblock Solving pomdps by searching the space of finite policies.
\newblock In {\em Proc. Fifteenth Conf. Uncertainty Artif. Intel.}, pages
  417--426. Morgan Kaufmann Publishers Inc., 1999.

\bibitem{doshi2015bayesian}
F.~Doshi-Velez, D.~Pfau, F.~Wood, and N.~Roy.
\newblock Bayesian nonparametric methods for partially-observable reinforcement
  learning.
\newblock {\em IEEE Trans. Patt. Anal. Mach. Intel.}, 37(2):394--407, 2015.

\bibitem{hausknecht2015deep}
M.~Hausknecht and P.~Stone.
\newblock Deep recurrent q-learning for partially observable {MDPS}.
\newblock {\em CoRR, abs/1507.06527}, 2015.

\end{thebibliography}
\end{document}